\documentclass[useAMS,usenatbib]{mn2e}

\title[Red clump stars of the Milky Way]
      {Red clump stars of the Milky Way -- laboratories of extra-mixing}

\author[G. Tautvai\v sien\.{e} et al.]
    { G. Tautvai\v sien\. e,$^{1}$\thanks{E-mail:grazina.tautvaisiene@tfai.vu.lt} 
G. Barisevi\v{c}ius,$^{1}$ Y. Chorniy,$^{1}$ I. Ilyin,$^{2}$ and E. Puzeras$^{1}$\\
       $^{1}$Institute of Theoretical Physics and Astronomy, Vilnius University, Go\v{s}tauto 
12, Vilnius 01108, Lithuania\\
       $^{2}$Leibniz Institut f\"{u}r Astrophysik Potsdam, An der Sternwarte 16, Potsdam 14482, Germany }

\begin{document}

\date{Accepted 2012 December 18. Received 2012 December 18; in original form 2012 November 26}

\pagerange{\pageref{firstpage}--\pageref{lastpage}} \pubyear{2013}

\maketitle

\label{firstpage}

\begin{abstract}

In this work we present the main atmospheric parameters, carbon, nitrogen and oxygen abundances, 
and $^{12}{\rm C}/^{13}{\rm C}$ ratios determined in a sample of 28 Galactic clump stars. 
Abundances of carbon were studied using the ${\rm C}_2$ 
band at 5086.2~{\AA}.  The wavelength interval 
7980--8130~{\AA} with strong CN features was analysed in order to determine 
nitrogen abundances and $^{12}{\rm C}/^{13}{\rm C}$  isotope ratios. 
The oxygen abundances were determined from the [O\,{\sc i}] line at 6300~{\AA}. 
The mean abundances of C, N and O abundances in the investigated clump stars support our previous 
estimations that, compared to the Sun and dwarf stars of the Galactic disk, carbon is depleted by 
about 0.2~dex, nitrogen is enhanced by 0.2~dex and oxygen is close to abundances in dwarfs.  
The $^{12}{\rm C}/^{13}{\rm C}$ and C/N ratios for galactic red clump stars analysed were compared to 
the evolutionary models of extra-mixing. The steeper drop of $^{12}{\rm C}/^{13}{\rm C}$ ratio in the 
model of thermohaline mixing by Charbonnel \& Lagarde better reflects the observational data at low stellar 
masses than the more shallow model of cool bottom processing  by Boothroyd \& Sackman. 
For stars of about $2~M_{\odot}$ masses a modelling of rotationally induced mixing should be 
considered with rotation of about 250\,km\,s$^{-1}$ at the time when a star was at the 
hydrogen-core-burning stage. 

\end{abstract}


\section{Introduction}

Investigations of galactic red clump stars may help 
to find fundamental answers to questions concerning mechanisms of transport of processed material to the stellar surface in 
low mass stars. Post-main-sequence stars with masses below $2-2.5~M_{\odot}$ exhibit signatures 
of material mixing that require challenging modelling beyond the standard stellar theory.
Carbon and nitrogen abundances are among most useful quantitative indicators of 
mixing processes in evolved stars. Already Iben (1965) has evaluated that the first dredge-up 
lowers abundances of $^{12}{\rm C}$ by about 30 per cent and abundances of $^{14}{\rm N}$ 
increase by about 80 per cent. Later on it was noticed that atmospheric abundances of stars are 
affected not only by the first dredge-up, but also by extra-mixing (Day, Lambert \& Sneden 1973; Pagel  1974; 
Tomkin, Luck \& Lambert 1976; Lambert \& Ries 1981; Gilroy 1989; Gilroy \& Brown 1991; 
Tautvai\v{s}ien\.{e} et al.\ 2000, 2001, 2005; Mikolaitis et al.\ 2010, 2011a,b, 2012, and references therein).

The extra-mixing becomes efficient at once 
after the luminosity function bump of the giant branch, i.e. the evolutionary point where the hydrogen-burning 
shell crosses the chemical discontinuity created by the outward moving convective envelope 
(Charbonnel 1994; Charbonnel, Brown \& Wallerstein 1998). The alterations depend on stellar mass, metallicity and evolutionary 
stage (e.g. Lambert \& Ries 1977; Boothroyd \& Sackmann 1999; Gratton et al.\ 2000, Spite et al.\ 2006; Charbonnel \& Zahn 2007; 
Recio-Blanco \& de Laverny 2007;  Tautvai\v{s}ien\.{e} et al.\ 2007, 2010; Smiljanic et al.\ 2009). 

Nowadays various mechanisms of extra-mixing were proposed by a number of scientific groups  
(see reviews by Chanam\'{e}, Pinsonneault \& Terndrup 2005; Charbonnel 2006; and recent papers by Cantiello \& Langer 2010; 
Denissenkov 2010; Lagarde et al.\ 2011; Palmerini et al.\ 2011; Wachlin, Miller Bertolami \& Althaus 2011; Angelou et al.\ 2012).

The $^{12}{\rm C/}^{13}{\rm C}$ ratio is the most robust diagnostic of deep mixing, because it 
is very sensitive to mixing processes and is almost insensitive to the adopted stellar parameters. 
We aim to study abundances of carbon isotopes, nitrogen and oxygen in an accurate and homogeneous way 
for a significant sample of clump star of the Galaxy. 

The {\it Hipparcos} catalogue (Perryman et al.\ 1997) contains about 600 clump stars with 
parallax errors lower than 10 per cent and represents a complete sample of clump stars to a distance 
of about 125~pc. Almost a half of these stars are already investigated by means of high resolution 
spectroscopy (Tautvai\v{s}ien\.{e} et al.\ 2003; 
Mishenina et al.\ 2006; Liu et al.\ 2007; Luck \& Heiter 2007; Tautvai\v{s}ien\.{e} \& Puzeras 2009; 
Puzeras et al.\ 2010); however, for the determination of carbon isotope ratios so far only our 
Paper~I (Tautvai\v{s}ien\.{e} et al.\ 2010) was dedicated.  

In Paper~I, we have determined  $^{12}$C, $^{13}$C, N and O abundances in atmospheres of 34 Galactic red 
clump stars. In this study the same data are determined for other 28 stars. 
The results of  $^{12}{\rm C/}^{13}{\rm C}$ and C/N ratios obtained in Paper~I and this work for the clump stars 
are compared with the available theoretical models of extra-mixing.

As concerns the red clump of our Galaxy, it was always interesting 
to find out what fraction of stars in the Galactic clump are first ascent giants and how many of them are  
He-core-burning stars (e.g. Mishenina et al.\ 2006; Paper~I). As it was shown in Paper~I, carbon isotope ratios 
allow us to obtain this information as well.

\section{Observational data and method of analysis}

Spectra were obtained at the Nordic Optical Telescope (NOT, La Palma) in 2006 December  
with the SOFIN \'{e}chelle spectrograph (Tuominen, Ilyin \& Petrov 1999).  
The second optical camera ($R\approx 80\,000$) was used to observe simultaneously 13 spectral orders, 
each of $40-60$~{\AA} in length, located from 5650~{\AA} to 8130~{\AA}. Two observations have been carried out  
for every star. A CCD frame was centered on carbon and nitrogen features during the first observation, 
and during the second  on the forbidden oxygen line. 
The reduction of CCD images was done using the {\sc 4A} software package 
(Ilyin 2000). Procedures of bias subtraction, 
cosmic ray removal, flat field 
correction, scattered light subtraction, and extraction of spectral orders were 
used for image processing. A Th-Ar comparison spectrum was used for the 
wavelength calibration. The continuum was defined from a number of narrow 
spectral regions, selected to be free of lines.

The spectra were analysed using a differential model atmosphere technique. 
 The {\sc Eqwidth} and {\sc Bsyn} program packages, developed at the Uppsala Astronomical 
  Observatory, were used to carry out the calculation of abundances from measured 
  equivalent widths and synthetic spectra, respectively. 
 A set of plane parallel, line-blanketed, constant-flux LTE model atmospheres 
was computed with an updated version of the {\sc MARCS} code (Gustafsson et al.\ 2008).

\input epsf
\begin{figure}
\epsfxsize=\hsize 
\epsfbox[5 10 385 370]{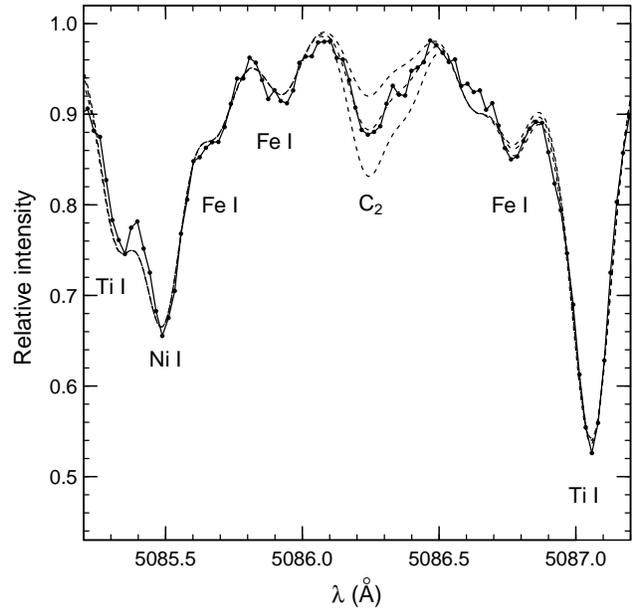} 
\caption{Synthetic and observed spectra for the C$_2$ region around $\lambda 5086$~\AA\ of HD\,417.  
The solid line shows the observed spectrum and the dashed lines show the synthetic 
spectra generated with ${\rm [C/Fe]}=-0.33$, $-0.23$ and $-0.13$.} 
\label{fig1}
\end{figure}

\input epsf
\begin{figure}
\epsfxsize=\hsize 
\epsfbox[5 10 495 330]{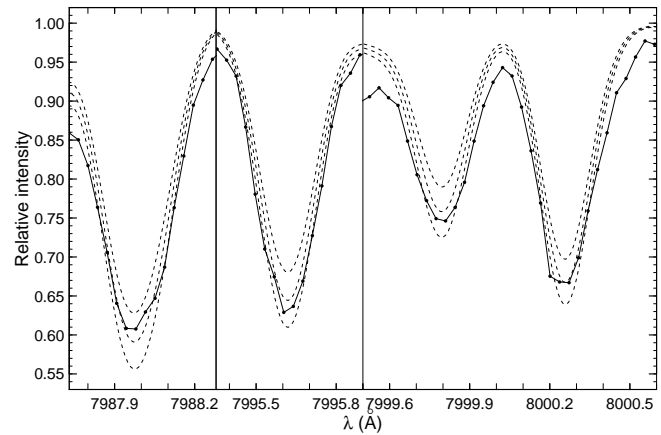} 
\caption{Stellar spectrum synthesis example around CN lines in HD\,6. The solid line shows the observed 
spectrum and the dashed lines show the synthetic spectra generated with ${\rm [N/Fe]}= 0.12$, 0.22 and 0.32.} 
\label{fig2}
\end{figure}

\input epsf
\begin{figure}
\epsfxsize=\hsize 
\epsfbox[5 10 435 370]{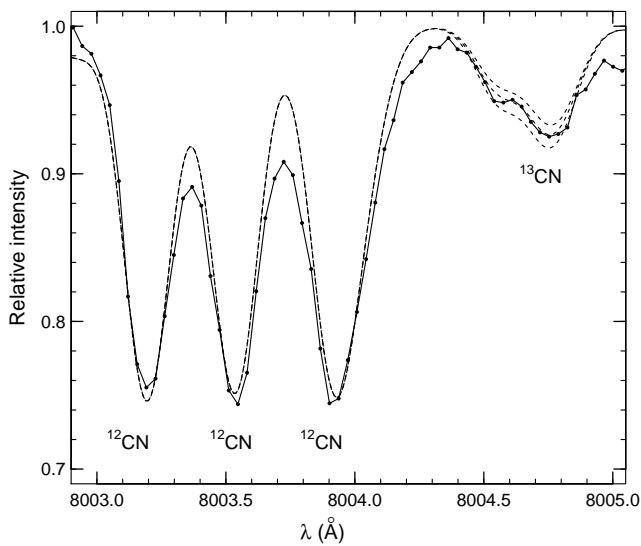} 
\caption{Stellar spectrum synthesis example around CN lines in the first ascent giant HD\,6. The solid line 
shows the observed spectrum and the dashed lines show the synthetic spectra generated with 
$^{12}{\rm C}/^{13}{\rm C}$ equal to 22 (upper line), 19 (middle line), and 16 (lower line).} 
\label{fig3}
\end{figure}

\input epsf
\begin{figure}
\epsfxsize=\hsize 
\epsfbox[5 10 435 370]{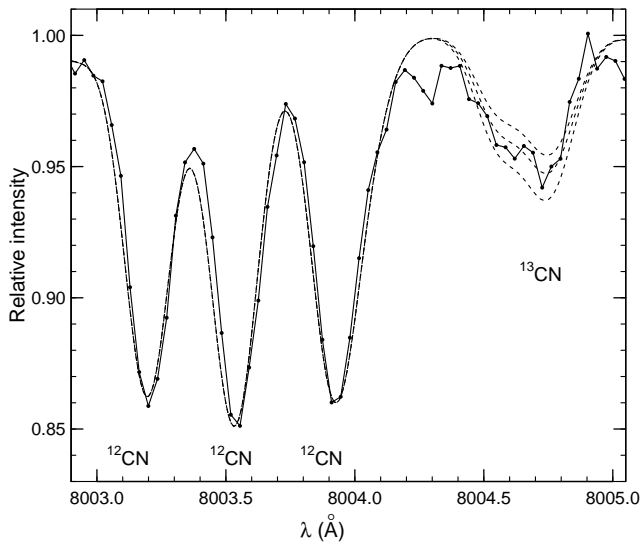} 
\caption{Stellar spectrum synthesis example around CN lines in the helium-core-burning star HD\,55730. 
The solid line shows the observed spectrum and the dashed lines show the synthetic spectra generated with 
$^{12}{\rm C}/^{13}{\rm C}$ equal to 12 (upper line), 10 (middle line), and 8 (lower line).} 
\label{fig4}
\end{figure}

Effective temperature, gravity and microturbulence were derived  using  
traditional spectroscopic criteria. The preliminary effective temperatures for 
the stars were determined using the $(B-V)_o$ and $(b-y)_o$ colour indices and the 
temperature calibrations by Alonso, Arribas \& Mart\'{i}nez-Roger (1999). 
All the effective temperatures were carefully checked and corrected, if needed, by 
forcing Fe~{\sc i} lines to yield no dependency of iron abundance on excitation 
potential by changing the model effective temperature. Surface gravity was obtained by 
forcing Fe~{\sc i} and Fe~{\sc ii} lines to yield the same 
[Fe/H]\footnote{In this paper we use the customary spectroscopic notation
[X/Y]$\equiv \log_{10}(N_{\rm X}/N_{\rm Y})_{\rm star} -\log_{10}(N_{\rm X}/N_{\rm Y})_\odot$}
value by 
adjusting the model gravity. The microturbulence value corresponding to minimal line-to-line  
Fe~{\sc i} abundance scattering was chosen as the correct value. 
Equivalent widths of 25--27 Fe~{\sc i} and 9--12 Fe~{\sc ii} lines were used for the determination of 
atmospheric parameters. The equivalent widths of the lines were measured by fitting 
a Gaussian profile using the {\sc 4A} software package (Ilyin 2000).     
Using the $gf$ values and solar equivalent widths of analysed iron lines from 
Gurtovenko \& Kostik (1989), we obtained the solar iron abundance, later used for the 
differential determination of iron abundances in the programme stars. We used the 
solar model atmosphere from the set calculated in Uppsala with a microturbulent 
velocity of 0.8~$\rm {km\,s}^{-1}$, as derived from Fe~{\sc i} lines. 

Carbon abundances were determined in stars using the 5085 -- 5087~{\AA} interval  
with the ${\rm C}_2$ band at 5086.2~{\AA}.  
The band at 5635.5~{\AA}, which we used in Paper~I, this time appeared at the very edge 
or outside the CCD frame, so following studies by Lambert \& Ries (1981) and Gratton et al.\ (2006) 
we have decided to analyse ${\rm C}_2$ $\lambda$~5086.2~{\AA}. We have checked the carbon abundance 
results given by both bands on HD~225216 and were satisfied by the agreement.
The interval 7980 -- 8130~{\AA} contains strong $^{12}{\rm C}^{14}{\rm N}$ and $^{13}{\rm C}^{14}{\rm N}$ 
features, so it was used for the nitrogen abundance and $^{12}{\rm C}/^{13}{\rm C}$ ratio analysis. 
The well known $^{13}{\rm CN}$ line at 8004.7~\AA\ was analysed in order to determine $^{12}{\rm C}/^{13}{\rm C}$ ratios. 
The molecular data for ${\rm C}_2$ and CN bands were provided by colleagues from the Uppsala Astronomical Observatory and 
slightly adjusted to fit the solar spectrum (Kurucz 2005) with the solar abundances by Grevesse \& Sauval (2000).
We present several examples of spectral syntheses of ${\rm C}_2$ and CN features and comparisons 
to the observed spectra in Figs.~1 -- 4.  

We derived oxygen abundances from the synthesis of the forbidden [O\,{\sc i}] line at 6300~{\AA} (Fig.~5 and 6). 
The $gf$ values for $^{58}{\rm Ni}$ and $^{60}{\rm Ni}$ isotopic line components, which blend the 
oxygen line, were taken from Johansson et al.\  (2003) and [O\,{\sc i}]  
log~$gf = -9.917$ value, as calibrated to the solar spectrum (Kurucz 2005).

The atomic oscillator strengths for stronger lines of iron and other elements were taken from 
Gurtovenko \& Kostik (1989). The Vienna Atomic Line Data Base (VALD, Piskunov et al.\ 1995) was extensively 
used in preparing the input data for the calculations. 
In addition to thermal and microturbulent Doppler broadening of lines, atomic 
line broadening by radiation damping and van der Waals damping were considered 
in the calculation of abundances. Radiation damping parameters of 
lines were taken from VALD. 
In most cases the hydrogen pressure damping of metal lines was treated using 
the modern quantum mechanical calculations by Anstee \& O'Mara (1995), 
Barklem \& O'Mara (1997) and Barklem, O'Mara \& Ross (1998). 
When using the Uns\"{o}ld (1955) approximation, correction factors to the classical 
van der Waals damping approximation by widths 
$(\Gamma_6)$ were taken from Simmons \& Blackwell (1982). For all other species a correction factor 
of 2.5 was applied to the classical $\Gamma_6$ $(\Delta {\rm log}C_{6}=+1.0$), 
following M\"{a}ckle et al.\ (1975). For lines stronger than 100~m{\AA} the correction factors were selected 
individually by inspection of the solar spectrum.

Stellar rotation was taken into account when it had noticeable influence to the profiles of lines. 
For HD\,26076, HD\,28191, HD\,38645, HD\,50371, 
and HD\,83805 the values of $v {\rm sin} i$ were taken from De Medeiros \& Mayor (1999); 
for the star HD\,225216  from Hekker \& Mel\'{e}ndez (2007); for HD\,5772  from Jones et al.\ (2011); 
and for HD\,35062, HD\,83805, HD\,223170, and HD\,224533 the values of  $v {\rm sin} i$ were adjusted based 
on the observed spectra in this work 
(1~km\,s$^{-1}$ for  HD\,35062, and 2~km\,s$^{-1}$ for the remaining three stars). 

Determinations of stellar masses were performed using effective temperatures, luminosities 
and isochrones of Girardi et al.\ (2000). The luminosities were calculated from {\it Hipparcos}  
parallaxes (van Leeuwen 2007), $V$ magnitudes were taken from SIMBAD, bolometric corrections calculated according to 
Alonso et al.\ (1999) and interstellar reddening corrections calculated using the Hakkila et al.\ (1994) software.

\subsection{Estimation of uncertainties}

\input epsf
\begin{figure}
\epsfxsize=\hsize 
\epsfbox[5 10 435 370]{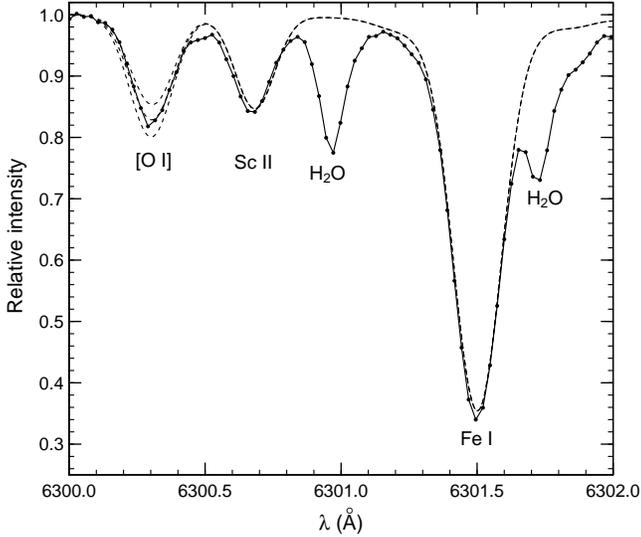} 
\caption{Stellar spectrum example in the region of [O\,{\sc i}] line $\lambda 6300.3$~\AA\ 
for HD\, 223170 (solid line) and the synthetic spectra calculated with [O/Fe]$= -0.13, -0.23$, and $-0.33$ 
(dashed lines).} 
\label{fig5}
\end{figure}

\input epsf
\begin{figure}
\epsfxsize=\hsize 
\epsfbox[5 10 435 370]{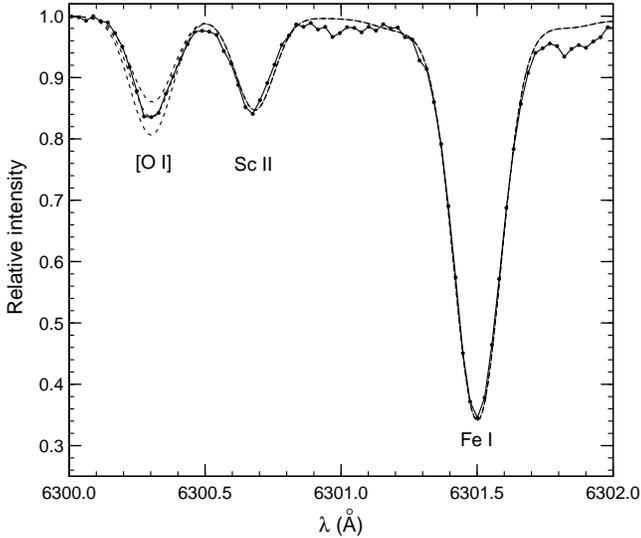} 
\caption{Stellar spectrum example in the region of [O\,{\sc i}] line $\lambda 6300.3$~\AA\ 
for HD\, 225216 (solid line) and the synthetic spectra calculated with [O/Fe]$= 0.06, -0.04$, and $-0.14$ 
(dashed lines).} 
\label{fig6}
\end{figure}

An approximate estimate of uncertainties due to random errors of the analysis can be evaluated from 
the line-to-line abundance scatter which is available for the Fe\,I, Fe\,II and nitrogen abundance determinations. 
The mean scatter of abundances is equal to 0.04~dex. 

The sensitivity of the abundance 
estimates to changes in the atmospheric parameters by the assumed errors is 
presented  for the stars HD\,6 and HD\,55730 in Table~1. 

Since abundances of C, N and O are bound together by the molecular equilibrium 
in the stellar atmosphere, we have also investigated how an error in one of 
them typically affects the abundance determination of another. 
In case of HD\,6: 
$\Delta{\rm [O/H]}=0.10$ causes 
$\Delta{\rm [C/H]}=0.03$ and $\Delta{\rm [N/H]}=0.10$;  
$\Delta{\rm [C/H]}=0.10$ causes $\Delta{\rm [N/H]}=-0.13$ and 
$\Delta{\rm [O/H]}=0.03$. $\Delta {\rm [N/H]}=0.10$ has no effect
on either the carbon or the oxygen abundances.
In case of HD\,55730: 
$\Delta{\rm [O/H]}=0.10$ causes 
$\Delta{\rm [C/H]}=0.03$ and $\Delta{\rm [N/H]}=0.06$;  
$\Delta{\rm [C/H]}=0.10$ causes $\Delta{\rm [N/H]}=-0.11$ and 
$\Delta{\rm [O/H]}=0.02$. $\Delta {\rm [N/H]}=0.10$ has no effect
on either the carbon or the oxygen abundances.

   \begin{table}
      \caption{Effects on derived abundances resulting from model changes. 
      The table entries show the effects on the 
logarithmic abundances relative to hydrogen, $\Delta$[A/H] } 
      \[
         \begin{tabular}{llrrc}
            \hline
            \noalign{\smallskip}
Star & Species & ${ \Delta T_{\rm eff} }\atop{ +100 {\rm~K} }$ & 
            ${ \Delta \log g }\atop{ +0.3 }$ & 
            ${ \Delta v_{\rm t} }\atop{ +0.3 {\rm km~s}^{-1}}$ \\
            \noalign{\smallskip}
            \hline
            \noalign{\smallskip}
HD 6 & & &  & \\
 &  C\,(C$_2$)      & 0.01 &  0.03 &  0.00 \\
 &  N\,(CN)         & 0.06 &  0.01 &  0.00 \\
 &  O\,([O{\sc i}]) &--0.01 &--0.05 &--0.01 \\  
 &  C/N             &--0.19  &--0.17 &  0.00 \\
 & $^{12}{\rm C}/^{13}{\rm C}$ &  0 & --2 & --2 \\
HD 55730 & & & & \\
 &  C\,(C$_2$)      & 0.02 &  0.03 &  0.00 \\
 &  N\,(CN)         & 0.10 &  0.01 &  0.00 \\
 &  O\,([O{\sc i}]) & 0.01 &--0.05 &--0.01 \\  
 &  C/N             &--0.26 & --0.10 &  0.00 \\
 & $^{12}{\rm C}/^{13}{\rm C}$ &  +2 & 0 & 0 \\  
                 \noalign{\smallskip}
            \hline
         \end{tabular}
      \]
   \end{table}


\section{Results and discussion}

The atmospheric parameters, abundances relative to hydrogen
[El/H], stellar masses and evolutionary stages 
determined for the programme stars are listed in Table~2. 
The evolutionary status to a star was attributed from its 
position on stellar evolutionary sequences in the luminosity versus effective 
temperature diagram by Girardi et al.\ (2000). HD\,39099 is marked by an asterisk 
since is a first ascent 
giant which has already past the luminosity bump of giant branch and though 
its $^{12}{\rm C}/^{13}{\rm C}$ isotope ratio is already lowered by extra-mixing. 

\begin{table*}
\centering
\caption{Atmospheric parameters and chemical element abundances of the programme stars}
\begin{tabular}{rcccrccrrrccccc}
\hline
HD & $T_{\rm eff}$ & log~$g$ & $v_{t}$ & [Fe/H] & $\sigma_{\rm Fe I}$ & $\sigma_{\rm Fe II}$  & [C/H] & [O/H] &[N/H] & $\sigma$  & C/N & $^{12}{\rm C}/^{13}{\rm C}$ & Mass & Ev.\\
   &    K   &     &  km\,s$^{-1}$    & & & 	&	&  &	&      &  &	 &	$M_\odot$  &   \\
\hline
6 & 4690 & 1.8 & 1.2 & $-0.03$ & 0.03 & 0.05 &$-0.26$ &  $-0.19$ & 0.14 & 0.05 & 1.41 & 19 & 1.4 & c  \\
417 & 4910 & 2.5 & 1.2 & $-0.13$ & 0.03 & 0.03 & $-0.36$ & $-0.07$ & 0.12 & 0.05 & 1.32 & 9 & 1.6 & c \\
448 & 4900 & 2.6 & 1.2 & 0.17 & 0.04 & 0.04 & $-0.10$ & 0.18 & 0.47 & 0.08 & 1.07 & 14 & 2.0 & c \\
756 & 4800 & 2.1 & 1.15 & $-0.18$ & 0.04 & 0.03 & $-0.47$ &$-0.29$ &  $-0.03$ & 0.06 & 1.45 & 13 & 1.5 & c \\
5722 & 4900 & 2.2 & 1.1 & $-0.19$ & 0.03 & 0.03 & $-0.51$ & $-0.21$ & 0.06 & 0.11 & 1.07 & 10 & 1.5 & c \\
9057 & 4870 & 2.3 & 1.3 & 0.02 & 0.04 & 0.03 & $-0.19$ & $-0.08$ & 0.33 & 0.06 & 1.20 & 19 & 2.5 & g \\
18690 & 4680 & 2.2 & 1.2 & 0.10 & 0.04 & 0.05 & $-0.13$ & $-0.02$ & 0.35 & 0.10 & 1.32 & 15 & 1.8 & c \\
21530 & 4630 & 2.1 & 1.1 & 0.17 & 0.04 & 0.04 & $-0.16$ & $-0.04$ & 0.31 & 0.04 & 1.35 & 18 & 1.4 & c \\
26162 & 4750 & 2.5 & 1.2 & 0.12 & 0.05 & 0.04 & $-0.07$ & 0.10 & 0.45 & 0.06 & 1.20 & 18 & 1.9 & c \\
26076 & 4810 & 2.1 & 1.2 & $-0.06$ & 0.03 & 0.05 & $-0.39$ & $-0.18$ & 0.12 & 0.06 & 1.23 & 11 & 1.9 & c \\
28191 & 4660 & 2.1 & 1.2 & 0.07 & 0.04 & 0.07 & $-0.21$ & $-0.13$ & 0.32 & 0.04 & 1.17 & 24 & 1.6 & g  \\
28625  & 4870 & 2.5 & 1.2 & 0.07 & 0.04 & 0.03 & $-0.21$ & $-0.01$ & 0.42 & 0.06 & 0.93 & 24 & 2.4 & g \\
31757  & 4740 & 2.0 & 1.1 & 0.17 & 0.03 & 0.04 & $-0.19$ & $-0.09$ & 0.32 & 0.07 & 1.23 & 22 & 2.6 & g \\
35062 & 4950 & 2.3 & 1.2 & 0.17 & 0.03 & 0.04 & $-0.23$ & $-0.08$ & 0.43 & 0.05 & 0.87 & 24 & 2.6 & g \\
39099  & 4650 & 2.2 & 1.2 & $-0.52$ & 0.03 & 0.04 & $-$ & $-$ &$-$ & $-$ & $-$ & 9 & 1.1 & g$^*$ \\
38645  & 4910 & 2.0 & 1.1 & $-0.06$ & 0.03 & 0.05 & $-0.42$ & $-0.35$ & 0.15 & 0.07 & 1.07 & 18 & 1.6 & c \\
40460  & 4830 & 2.2 & 1.1 & $-0.13$ & 0.03 & 0.05 & $-0.32$ & $-0.11$ & 0.06 & 0.03 & 1.66 & $-$ & 1.6 & c \\
41125 & 4930 & 2.4 & 1.1 & $-0.07$ & 0.04 & 0.05 & $-0.39$ & $-0.15$ & 0.10 & 0.05 & 1.29 & 24 & 2.2 & g \\
41783  & 4820 & 2.0 & 1.1 & $-0.15$ & 0.02 & 0.05 & $-0.58$ & $-0.37$ & 0.03 & 0.03 & 0.98 & 16 & 1.5 & c \\
43043 & 4860 & 2.1 & 1.45 & $-0.30$ & 0.02 & 0.05 & $-0.64$ & $-0.36$ & $-0.06$ & 0.04 & 1.05 & 19 & 1.4 & c \\
44061 & 4760 & 2.5 & 1.2 & $-0.18$ & 0.03 & 0.04 & $-0.35$ & 0.00 & 0.21 & 0.05 & 1.10 & 24 & 1.9 & g \\
50371 & 4980 & 2.5 & 1.2 & 0.06 & 0.03 & 0.04 & $-$ & $-$ & $-$ & $-$ & $-$ & 15 & 2.3 & c \\
55730 & 4830 & 2.1 & 1.2 & $-0.14$ & 0.03 & 0.06 & $-0.43$ & $-0.26$ & 0.07 & 0.04 & 1.26 & 10 & 1.4 & c \\
63410  & 4860 & 2.0 & 1.05 & $-0.27$ & 0.03 & 0.04 & $-0.51$ & $-0.32$ & $-0.17$ & 0.06 & 1.82 & 8 & 1.2 & c \\
83805 & 4930 & 2.1 & 1.1 & 0.00 & 0.04 & 0.04 & $-0.37$ & $-0.23$ & 0.35 & 0.08 & 0.76 & 9 & 2.2 & c \\
223170  & 4760 & 2.5 & 1.1 & 0.15 & 0.04 & 0.04 & $-0.08$ & 0.17 & 0.37 & 0.03 & 1.41 & 45 & 2.1 & g \\
224533  & 4950 & 2.4 & 1.1 & 0.01 & 0.04 & 0.05 & $-0.34$ & $-0.16$ & 0.24 & 0.07 & 1.05 & 21 & 2.5 & g \\
225216 &  4680 & 2.3 & 1.15 & $-0.02$ & 0.04 & 0.03 & $-0.25$ & $-0.06$ & 0.16 & 0.02 & 1.55 & 14 & 1.5 & c \\
\hline
\end{tabular}
\medskip
Evol.: g -- first ascent giant, c -- He-core-burning star,  $^*$ -- first ascent giant above the bump.
\end{table*}

Our sample of investigated stars almost does not overlap with samples of clump stars investigated in other 
recent studies. Only one star is in common with the sample by Mishenina et al.\ (2006), one with Liu et al.\ (2007), 
and 8 stars are in common with Luck \& Heiter (2007). As well as in the comparison which we have made between 
our paper Puzeras et al.\ (2010) and Luck \& Heiter (then we had 16 common stars), we do not find obvious systematic differences 
between the main atmospheric parameters determined in these studies. Luck \& Heiter have provided spectroscopic and physical 
atmospheric parameters. When compared to the physical parameters, [Fe/H]$_{\rm Our} - {\rm [Fe/H]}_{\rm LH} = +0.08 \pm 0.06$ (24 stars), 
$T_{\rm eff(Our)} - T_{\rm eff(LH)} = -1 \pm 66$, and log~$g_{\rm Our} - {\rm log}~g_{\rm LH} = -0.15 \pm 0.15$. 
In case of spectroscopic parameters, [Fe/H]$_{\rm Our} - {\rm [Fe/H]}_{\rm LH} = 0.00 \pm 0.06$, 
$T_{\rm eff(Our)} - T_{\rm eff(LH)} = -88 \pm 60$, and log~$g_{\rm Our} - {\rm log}~g_{\rm LH} = -0.44 \pm 0.25$. 
See Puzeras et al.\ (2010) for a discussion and comparisons between other studies. 

C, N and O abundances in the investigated stars are similar to the results of Paper~I 
and confirm the conclusion of Paper~I that compared with the Sun and dwarf 
stars of the Galactic disk, the mean abundance of carbon in clump stars 
is depleted by about 0.2~dex, nitrogen is enhanced by 0.2~dex and oxygen is 
close to abundances in dwarfs.

\input epsf
\begin{figure*}
\epsfxsize=22cm 
\epsfbox[0 5 630 280]{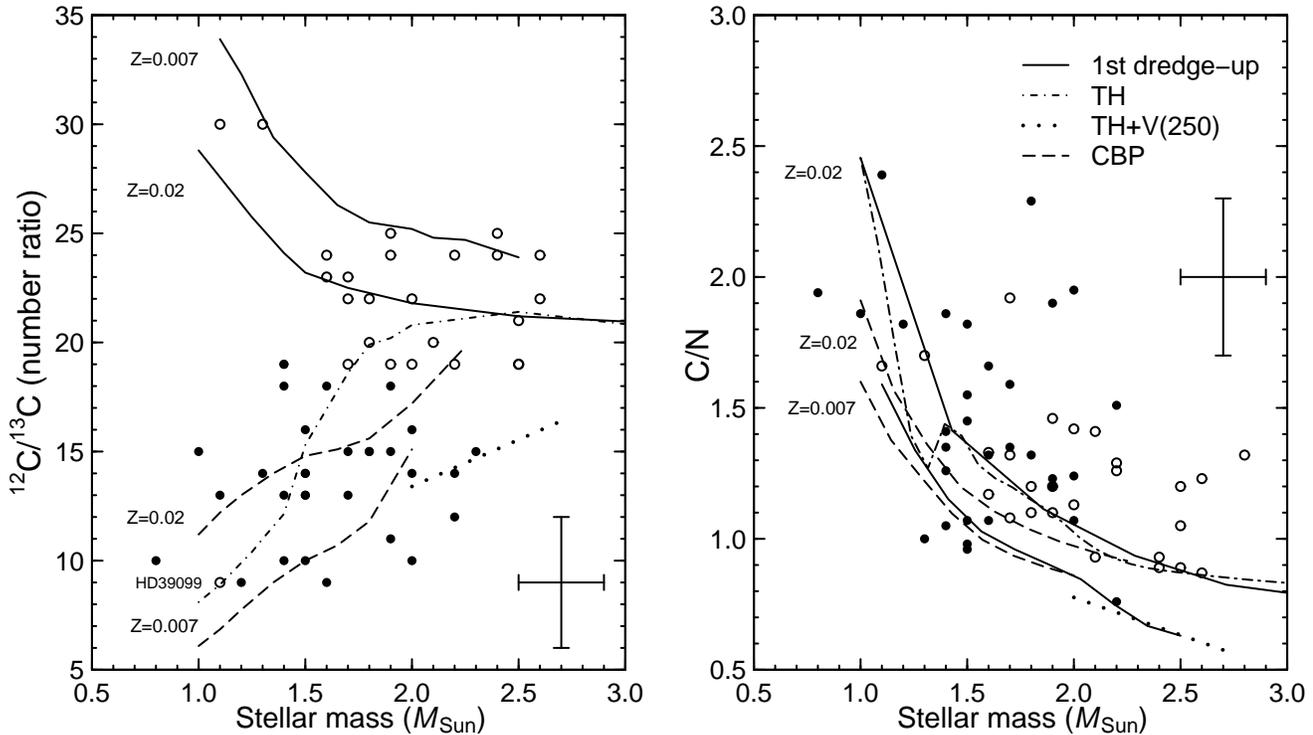} 
\caption{ C/N and $^{12}{\rm C}/^{13}{\rm C}$ ratios 
versus stellar turn-off mass for galactic red clump stars analysed in Paper~I and this work. The stars identified 
as first ascent giants are shown by empty circles, and the stars identified as helium-core-burning stars -- filled circles. 
The theoretical curves are taken from Boothroyd \& Sackmann (1999) -- solid and long dashed lines 
and Charbonnel \& Lagarde (2010) -- dashed-dotted and dotted lines. See Sect.~3.1 for more explanations.} 
\label{fig7}
\end{figure*}

\subsection{Comparisons with theoretical models of extra-mixing}

The determined $^{12}{\rm C}/^{13}{\rm C}$ and C/N ratios we compared with two theoretical models.

A recent so-called thermohaline model of extra-mixing provides evolutionary trends of 
$^{12}{\rm C}/^{13}{\rm C}$ and C/N ratios versus stellar masses for solar metallicity stars  
(Charbonnel \& Lagarde 2010). In order to develop this model of mixing, the combination of new ideas and 40-yr-old  
predictions by Ulrich (1972) was done. 
Eggleton et al.\ (2006) found a mean molecular weight ($\mu$) inversion in their $1 M_{\odot}$ stellar 
evolution model, occurring after the so-called luminosity bump on the red giant branch, when the H-burning shell 
source enters the chemically homogeneous part of the envelope. The $\mu$ inversion is produced by the reaction 
$^3{\rm He(}^3{\rm He,2p)}^4{\rm He}$, as predicted by Ulrich (1972). It does not occur earlier, because the 
magnitude of the $\mu$ inversion is small and negligible compared to a stabilizing $\mu$ stratification. 
Charbonnel \& Zahn (2007) also computed stellar models 
including the prescription by Ulrich (1972) and extend them to the case of a non-perfect gas for the turbulent 
diffusivity produced by that instability in stellar radiative zone. They found that a double diffusive instability 
referred to as thermohaline convection, which has been discussed long ago in the literature (Stern 1960), 
is important in the evolution of red giants. This mixing connects the convective envelope with the external wing of 
the hydrogen burning shell and induces surface abundance modifications in red giant stars (Charbonnel \& Lagarde 2010). 
Unfortunately, this model is computed so far only for solar metallicity stars. 

An other so called cool bottom processing model has been developed more than a decade ago  
(Boothroyd et al.\ 1995; Wasserburg, Boothroyd \& Sackman 1995; Boothroyd \& Sackman 1999, and references therein).
The model by Boothroyd \& Sackmann (1999) includes the deep circulation mixing below the base of 
the standard convective envelope, and the consequent "cool bottom processing" of CNO isotopes. In this model, 
an extra-mixing takes material from the convective envelope, transports it down to regions hot enough for some nuclear
 processing in the outer wing of H-burning shell, and then transports it back up to the convective envelope. For the 
 computations of the extra-mixing, the "conveyor-belt" circulation model was used. The temperature difference between 
 the bottom of mixing and the bottom of the H-burning shell was considered as a free parameter, to be determined by 
 comparison with observations. For this purpose, the authors used data of M~67 (Gilroy 1989; Gilroy \& Brown 1991). 

In Fig.~7, $^{12}{\rm C}/^{13}{\rm C}$ and C/N ratios versus stellar mass for galactic red clump stars 
analysed in Paper~I and in this work are compared with the mentioned theoretical models. The carbon isotope ratio versus 
stellar mass plot is much more informative than the plot of carbon-to-nitrogen ratios. In the upper part of the 
 $^{12}{\rm C}/^{13}{\rm C}$ ratio versus stellar mass plot, along to the first dredge-up sequences of two metallicities 
($Z=0.02$ and 0.007) the first ascent giants are located.  Carbon isotope ratios of these stars are altered according 
to the 1$^{st}$ dredge-up prediction. 

Helium-core-burning stars have $^{12}{\rm C}/^{13}{\rm C}$ ratios already lowered by extra mixing and lie in the 
lower part of the $^{12}{\rm C}/^{13}{\rm C}$ versus stellar mass diagram. It is seen that the trend of lowering of carbon 
isotope ratios in low-mass He-core-burning stars is quite steep. Carbon isotope ratios may be quite different in stars of 
quite similar masses, for example, in stars of about $1.5\,M_{\odot}$, the $^{12}{\rm C}/^{13}{\rm C}$ values range from 
about 18 to 9. The theoretical model of thermohaline mixing is better reflecting the observational data in this respect  
than the more shallow model of cool bottom processing. 

As concerns the helium-core-burning stars of higher masses, it seems that the mixing was more intense, and this might 
be caused by a stellar rotation while stars were in a stage of hydrogen burning in their cores.   
Charbonnel \& Lagarde (2010) computed models with rotation-induced mixing for stars at the zero age main 
sequence (ZAMS) having rotational velocities of 110, 250 and 300\,km\,s$^{-1}$. 
Typical initial ZAMS rotation
velocities were chosen depending on the stellar mass based on observed rotation distributions in young open clusters
(Gaige 1993). The convective envelope was supposed to rotate as a solid body through the evolution. The transport
coefficients for chemicals associated to thermohaline and rotation-induced mixing were simply added in the diffusion
equation and the possible interactions between the two mechanisms were not considered. Interesting to note that the 
rotation-induced mixing modifies the internal chemical structure of main sequence stars, although its signatures are 
revealed only later in the stellar evolution. 

In Fig.~7 we plotted a model corresponding to the rotational velocity of 250\,km\,s$^{-1}$.
The observational data for stars of about  $2~M_{\odot}$ in our sample require a modelling of mixing with rotation 
of about this speed when stars were on the main sequence. Even large rotationally induced mixing is suggested by stars 
of high masses in open clusters (e.g. Tautvai\v{s}ien\.{e} et al.\ 2005; Smiljanic et al.\ 2009; Mikolaitis et al.\ 2010, 2011b).  

\subsection{Membership of the Galactic red clump}

The stellar positions in the 
$^{12}{\rm C}/^{13}{\rm C}$ versus stellar mass diagram as well as comparisons to stellar evolutionary 
sequences in the luminosity versus effective temperature diagram may show which stars 
are helium-core-burning ones and which are first ascent giants. From the available sample of 62 Galactic 
red clump stars investigated in this work and Paper~I, 35 were identified as helium-core-burning ones and 27 
as first ascent just hydrogen-shell-burning giants.   
In the paper by Mishenina et al.\ (2006), according to nitrogen abundance values, 
the authors have suggested 21 helium-core-burning stars, about 54 candidates to helium-core-burning stars 
and about 100 first ascent giants. Looking to the density of stars above and below the clump in the 
colour-magnitude diagram for the {\it Hipparcos} catalogue, it seems that a number of helium-core-burning stars 
in the Galactic red clump should be slightly larger than a number of first ascent giants. For example, if we count 
a number of stars in a frame of $M_v = 2 \pm 0.5$ and $0.8 < B-V < 1.2$ targeted on the clump, we count 1050 
stars, while below the clump, in a frame of  $M_v = 1 \pm 0.5$ and $0.8 < B-V < 1.2$, we have 380 stars. 

A very promising way to distinguish between hydrogen- and helium-burning red giant stars was recently uncovered 
by asteroseismic observations (Bedding et al.\ 2011). From high-precision photometry obtained by the {\sc Kepler} space 
observatory, it was found that gravity-mode period spacings are different in  hydrogen- and helium-burning red giants. 
In hydrogen-shell-burning stars the period spacing is mostly about 50 seconds, and in helium-burning 
stars it is from about 100 to 300 seconds. These differences appear because of differences of internal structure 
and core densities of stars. This asteroseismic way of stellar evolutionary status determination is especially 
important for stars of larger than $2~M_{\odot}$ masses for which chemical composition 
signatures of extra-mixing are less trustful.

\subsection{Conclusions}

In this work we present the main atmospheric parameters, abundances of nitrogen, carbon and oxygen, 
and $^{12}{\rm C}/^{13}{\rm C}$ ratios for a sample of 28 Galactic clump 
stars from high-resolution spectra observed on the Nordic Optical telescope ($R\approx 80\,000$). 
 
The mean abundances of C, N and O in the investigated clump stars support our previous 
estimations (Paper~I) that compared to dwarf stars of the Galactic disk carbon is depleted by 
about 0.2~dex, nitrogen is enhanced by 0.2~dex and oxygen is close to abundances in dwarfs.  

The $^{12}{\rm C}/^{13}{\rm C}$ and C/N ratios for the galactic red clump stars analysed were compared to 
the evolutionary models of extra-mixing. The steeper drop of $^{12}{\rm C}/^{13}{\rm C}$ ratio in the model of 
thermohaline mixing (Charbonnel \& Lagarde 2010) better reflects the observational data at low stellar masses 
than the more shallow model of cool bottom processing (Boothroyd \& Sackman 1999). 

In our sample, for stars of about $2~M_{\odot}$ masses the modelling of rotationally induced mixing should be 
considered with rotation of about 250\,km\,s$^{-1}$ at the time when a star was at the hydrogen-core-burning stage. 

The stellar positions in the 
$^{12}{\rm C}/^{13}{\rm C}$ versus stellar mass diagram as well as comparisons to stellar evolutionary 
sequences in the luminosity versus effective temperature diagram by Girardi et al.\ (2000) 
show that from the investigated sample of Galactic red clump stars about a half are first ascent 
giants with carbon isotope ratios altered according to the first dredge-up prediction, and an other 
half are helium-core-burning stars with carbon isotope ratios altered by extra-mixing.

\section*{Acknowledgements}

We wish to thank sincerely B. Edvardsson, B. Gustaffson and K. Eriksson (Uppsala Observatory) for 
providing computing codes and atomic data for this work. 
This project has been supported by the European Commission through ``Access to 
Research Infrastructures Action" of the ``Improving Human Potential Programme", awarded 
to the Instituto de Astrof\' isica de Canarias to fund European Astronomers' access to the 
European Nordern Observatory, in the Canary Islands.

\bsp

\label{lastpage}

\end{document}